\documentclass[12pt]{iopart}

\usepackage{iopams}  
\usepackage{bm}
\usepackage{here}
\usepackage{graphicx}

\begin{document}

\title[Reversible superconducting-normal phase transition in a magnetic field]{Reversible superconducting-normal phase transition in a magnetic field: The energy-momentum balance
including the velocity field of the Berry connection from many-body wave functions}
\author{Hiroyasu Koizumi}

\address{Division of Quantum Condensed Matter Physics, Center for Computational Sciences, University of Tsukuba, Tsukuba, Ibaraki 305-8577, Japan}
\ead{koizumi.hiroyasu.fn@u.tsukuba.ac.jp}
\vspace{10pt}

\begin{abstract}
The velocity field composed of the Berry connection from many-body wave functions and electromagnetic vector potential explains the energy-momentum balance during the reversible superconducting-normal phase transition in the presence of an externally applied magnetic field. 
In this formalism, forces acting on electrons are the Lorentz force and force expressed as the gradient of the kinetic energy. In the stationary situation, they balance; however, an infinitesimal imbalance of them causes a phase boundary shift. 
In order to explain the energy balance during this phase boundary shift, the electromotive force of the Faraday's magnetic induction type is considered for the Berry connection. 
This theory assumes that supercurrent exists as a collection of stable quantized loop currents, and the transition from the superconducting to normal phase is due to the loss of their stabilizations through the thermal 
fluctuation of the winding numbers of the loop currents. We argue that an abrupt change of loop current states with integral quantum numbers should be treated as a quantum transition; then, the direct conversion of the quantized loop currents to the magnetic field occurs; consequently, the Joule heat generation does not occur during the phase transition.
\end{abstract}

%
%
%
%
%

\section{Introduction}

There have been controversies about the reversibility of the superconducting-normal phase transition in the presence of a magnetic field \cite{Hirsch2017,Hirsch2018,Hirsch2020}. Experiments on type-I superconductors indicate that it is a thermodynamically reversible process without Joule heat generation \cite{Keesom1,Keesom2,Keesom3, Keesom4,Keesom}. 
However, its theoretical explanation within the standard theory of superconductivity has not been successful.
The most serious point is that the theory needs to take into account the electric field generation by the Faraday's law, 
\begin{eqnarray}
\nabla \times {\bf E} = -\partial_t {\bf B}
  \label{Faraday}
\end{eqnarray}
where ${\bf E}$ is the electric field and $ {\bf B}$ is the magnetic field. 
Since the superconducting to normal phase transition is accompanied by
the change of the magnetic field due to the fact that the magnetic field is zero in the superconducting phase, and non-zero in the normal phase. In the boundary of the superconducting and normal phases,
 the current called the Meissner current exists in the superconducting phase side, which is absent in the normal phase side.
Then, the disappearance of it needs to happen upon the change of the superconducting to normal phase. According to the standard theory for electric conduction, it should occur through the energy dissipation by the Joule heating 
\begin{eqnarray}
{\bf j}\cdot{\bf E},
\end{eqnarray}
where ${\bf j}$ is the current density. 
However, this generation of the Joule heat makes the phase transition irreversible, which disagrees with the experimental reversible transition.

In order to overcome the above difficulty, rethinking of Eq.~(\ref{Faraday}) has been proposed \cite{koizumi2020b}.
First, the formalism using 
 the vector potential ${\bf A}$ and scalar potential $\varphi$ is employed, where
${\bf E}$ and ${\bf B}$ are related to ${\bf A}$ and $\varphi$ as
\begin{eqnarray}
{\bf E}=-\partial_t {\bf A}-\nabla \varphi, \quad {\bf B}=\nabla \times {\bf A}
\label{eqEA}
\end{eqnarray}
Then, Eq.~(\ref{Faraday}) is automatically satisfied. 
The use of the vector potential instead of the electric and magnetic fields is in accordance with
the Aharonov-Bohm effect which indicates that ${\bf A}$ and $\varphi$ 
are physically more fundamental than
 ${\bf E}$ and ${\bf B}$ \cite{AB1959,Tonomura1982,Tonomura1986}.
Second, the vector potential that behaves in a similar manner as the electromagnetic vector potential, which was discovered by Berry \cite{Berry,Geometric} is added. We call this vector potential type object the ``Berry connection''.

Recently, a way to obtain a Berry connection in many-electron system has been developed. It is called the Berry connection from many-body wave functions, and its usefulness has been demonstrated \cite{koizumi2019,koizumi2022,koizumi2022b,Koizumi2023}. It has been argued that the vector potential formalism including both electromagnetic one and the Berry connection explains the energy conversion between the Meissner current kinetic energy and the magnetic field energy \cite{koizumi2020b,koizumi2022,koizumi2022b}.
 In the present work, we extends the above discuss. 

The organization of the present work is as follows: In Section~\ref{Sec2}, the velocity field for superconducting electrons 
is described using the electromagnetic vector potential and Berry connection from many-body wave functions.
From the time-derivative of the velocity field, the equation for the force balance is obtained. It includes a
contribution expressed as the gradient of the kinetic energy, whihc has been missing in the arguments so far \cite{Hirsch2017,Hirsch2018,Hirsch2020}. In Section~\ref{Sec3}, conditions for the local velocity field stationarity
under the existence of the fluctuation of the loop currents are presented. We argue that the electromotive force of the Faraday's magnetic induction type needs to be used instead of the simple Newtonian vector force balance.
It is also argued that an abrupt change of loop current states with integral quantum numbers should be treated as a quantum transition. In Section~\ref{Sec4}, we show that the reversible superconducting-normal phase transition in a magnetic field is possible thanks to the quantum transition of the loop current states that does not generate electric field.
In Section~\ref{Sec5}, we conclude the present work by mentioning  the similarity of the present work with
the Maxwell's vector potential based formalism for electromagnetic phenomena \cite{Maxwell1,Maxwell2,Maxwell3,Maxwell4}.

\section{The total time derivative of the velocity field and the force balance
in superconductors}
\label{Sec2}

We include the 
Berry connection from many-body wave functions defined by
\begin{eqnarray}
\! \!
\!{\bf A}^{\rm MB}_{\Psi}({\bf r},t)\!=\!
{{{\rm Re} \left\{
 \int d\sigma_1  d{\bf x}_{2}  \cdots d{\bf x}_{N}
 \Psi^{\ast}({\bf r}, \sigma_1, \cdots, {\bf x}_{N},t)
  (-i \hbar \nabla )
\Psi({\bf r}, \sigma_1, \cdots, {\bf x}_{N},t) \right\}
 }
 \over {\hbar \rho({\bf r},t)}} 
\nonumber
\\
\label{Afic}
\end{eqnarray}
in addition to the electromagnetic vector potential in this work.
Here,
`$\rm{Re}$' denotes the real part, $\Psi$ is the total electronic wave function, ${\bf x}_i$ collectively stands for the coordinate ${\bf r}_i$ and the spin $\sigma_i$ of the $i$th electron, $-i \hbar \nabla$ is the Schr\"{o}dinger's momentum operator for the coordinate vector ${\bf r}$, and $\rho({\bf r},t)$ is the number
density calculated from $\Psi$. This Berry connection is obtained by regarding ${\bf r}$ as the ``adiabatic parameter''\cite{Berry}. 

For convenience, we also use the following $\chi$ defined as
  \begin{eqnarray}
{ {\chi({\bf r},t)}}= - 2\int^{{\bf r}}_0 {\bf A}_{\Psi}^{\rm MB}({\bf r}',t) \cdot d{\bf r}' 
\end{eqnarray}
It is an angular variable with period $2\pi$ and 
\begin{eqnarray}
w_C[\chi]={1 \over {2\pi}}\oint_C \nabla \chi \cdot d{\bf r}
\label{winding}
\end{eqnarray}
 is the topological integer called, the `winding number'.  When singularities of $\chi$ exist within the loop $C$,
 $w_C[\chi]$ may become non-zero integer. The non-zero winding number case explains the flux quantization in the unit ${h \over {2e}}$ \cite{koizumi2022b}.

Using $\chi$, the many-electron wave function $\Psi$ is expressed as
\begin{eqnarray}
\Psi({\bf x}_1, \cdots, {\bf x}_N,t)=\exp \left(  -{i \over 2} \sum_{j=1}^{N} \chi({\bf r}_j,t)\right)
\Psi_0({\bf x}_1, \cdots, {\bf x}_N,t)
\label{single-valued}
\end{eqnarray}
with $\Psi_0$ being a currentless wave function \cite{koizumi2019,koizumi2022,koizumi2022b,Koizumi2023}.
The current density for $\Psi$ is given by
\begin{eqnarray}
{\bf j}=-e \rho {\bf v}
\label{eq11}
\end{eqnarray}
with ${\bf v}$ being the velocity field 
\begin{eqnarray}
{\bf v}={e \over m_e}{\bf A}+{\hbar \over {m_e}}{\bf A}_{\Psi}^{\rm MB}
={e \over m_e}\left({\bf A}-{\hbar \over {2e}}\nabla \chi \right)
\label{eq12}
\end{eqnarray}
Actually, the supercurrent electron density, $n_s$, is usually different from $\rho$ \cite{koizumi2022,koizumi2022b}.
Thus, we use the following formula
\begin{eqnarray}
{\bf j}=-e n_s{\bf v}
\label{supercurrentd}
\end{eqnarray}
for the supercurrent density in the following.

Let us consider the situation where a superconductor exists $x \ge 0$, and the magnetic field is applied in the $z$-direction. We assume that the $x <0$ region is the vacuum, and the magnetic field given by $B_0 {\bf e}_z$ exists there, where ${\bf e}_a$ denotes the unit vector in the $a$-direction.
In this situation, the magnetic field in the superconductor is given by
\begin{eqnarray}
{\bf B}=B_0 {\bf e}_z e^{ -x/\lambda_L}
\label{eq11}
\label{lambda}
\end{eqnarray}
where $\lambda_L$ is the London penetration depth
\begin{eqnarray}
\lambda_L= \sqrt{m_e \over {\mu_0 n_s e^2}}
\label{eq12L}
\end{eqnarray}
with $\mu_0$ being the vacuum permeability \cite{London1950}.
Due to the presence of the factor $e^{ -x/\lambda_L}$ in ${\bf B}$, the magnetic field is appreciable only in the surface region $0 <x < \lambda_L$.
Using the Amp\'{e}re's law $\nabla \times {\bf B}=\mu_0 {\bf j}$, the current density is calculated as
\begin{eqnarray}
{\bf j}={1 \over {\mu_0 \lambda_L}}B_0 {\bf e}_ ye^{ -x/\lambda_L}
\label{j1}
\end{eqnarray}
This is the Meissner current.
From this ${\bf j}$, the velocity field is obtained as
\begin{eqnarray}
{\bf v}=-{ 1 \over {n_s e}}{\bf j}=-{1 \over {n_s e\mu_0 \lambda_L}}B_0 {\bf e}_ ye^{ -x/\lambda_L}
\label{v1}
\end{eqnarray}

Now, we consider the total time-derivative of the velocity field. According to the Eulerian view of the time-derivative of a field, 
the total time-derivative of a field $f(x,y,z,t)$ is given by
\begin{eqnarray}
 {{d f} \over {dt}}= \partial_t{f}+({\bf v} \cdot \nabla){f}
\end{eqnarray}
Thus, the total time-derivative of ${\bf v}$ is given by
\begin{eqnarray}
{{d{\bf v}} \over {dt}}&=&\partial_t{\bf v}+({\bf v} \cdot \nabla){\bf v}
\nonumber
\\
&=&\partial_t{\bf v}+
{1 \over 2}\nabla v^2-{\bf v}\times(\nabla \times {\bf v})
\nonumber
\\
&=& {e \over m_e}\partial_t \left({\bf A}-{\hbar \over {2e}}\nabla \chi \right)+
{1 \over 2}\nabla v^2-{{e \over m_e}\bf v}\times \left[{\bf B}-{\hbar \over {2e}}\nabla \times (\nabla \chi) \right]
\label{eq13}
\end{eqnarray}
where the vector identify $({\bf v} \cdot \nabla){\bf v}={1 \over 2}\nabla v^2-{\bf v}\times(\nabla \times {\bf v})$ is used.

\begin{figure}[H]
\begin{center}
\includegraphics[width=8.0cm]{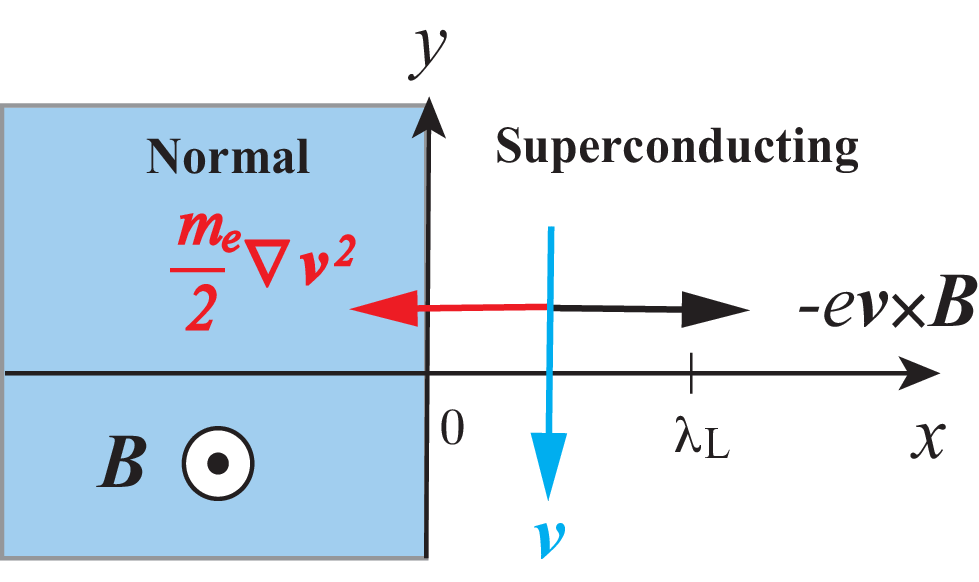}
\end{center}
\caption{A situation for the normal-superconducting phase transition in a magnetic field. The superconductor phase exists $x \ge 0$, and the normal phase 
exists in $x <  0$. The magnetic field in the normal phase region is ${\bf B}=B_0 {\bf e}_z$,
and that in the superconducting phase region is ${\bf B}=B_0 {\bf e}_z e^{ -x/\lambda_L}$.
  The Meissner current in the interface region is generated by the velocity field ${\bf v}=-(n_s e\mu_0 \lambda_L)^{-1}B_0 {\bf e}_ ye^{ -x/\lambda_L}$. 
  In the stationary situation, the Lorentz force $-e{\bf v} \times {\bf B}$ and the gradient of the kinetic energy force ${ m_e \over 2}\nabla v^2$ balance.}
\label{S-N-inter}
\end{figure}

Let us consider the superconducting-normal state transition case. 
For definiteness, we consider the situation where the superconductor phase exists in the $x \ge 0$ region, and the normal phase 
exists in the $x <  0$ region; the applied magnetic field is in the $z$-direction given by $B_0 {\bf e}_z$ (see Fig.~\ref{S-N-inter}). The magnetic field in the superconducting region is given by Eq.~(\ref{lambda}), and the supercurrent density by Eq.~(\ref{j1}). We obtain the following balance relation
\begin{eqnarray}
{m_e \over 2}\nabla v^2-e{\bf v}\times{\bf B}=0
\label{eqfb}
\end{eqnarray}
between the Lorentz force and the gradient of the kinetic energy force
from Eqs.~(\ref{eq11}),(\ref{eq12L}), and (\ref{v1}).
 Note that this gradient of the kinetic energy force is absent in the classical treatment \cite{Hirsch2017,Hirsch2018,Hirsch2020}, thus, makes it difficult to explain how the the Lorentz force is balanced.
 
Using Eq.~(\ref{eqfb}), the stationary condition from Eq.~(\ref{eq13}) becomes
\begin{eqnarray}
\partial_t{\bf A}={ \hbar \over {2e}} \left[ \partial_t (\nabla \chi)-{\bf v} \times \left(\nabla \times (\nabla \chi) \right)\right]
\label{cbfb}
\end{eqnarray}
We will examine this relation in the following.

\section{The energy balance for superconducting-normal phase transition in a magnetic field: Stationary case. }
\label{Sec3}

Let us consider the condition in Eq.~(\ref{cbfb}). It cannot be satisfied as it is. Instead, we consider the equality by integrating the both sides along a loop.
This change of the meaning of the equality is motivated by the fact that the Faraday's formula for the electromotive force generation by magnetic induction has such a form;
the electromotive force for an electron
\begin{eqnarray}
{\cal E}={1 \over {-e}}\oint_C {{d (m_e{\bf v})} \over {dt}}\cdot d{\bf r}
\label{eq21r}
\end{eqnarray}
and the electromotive force generation by magnetic induction
\begin{eqnarray}
{\cal E}=-{d \over {dt}}\int_S {\bf B}\cdot d{\bf S}
\label{eq21l}
\end{eqnarray}
are equated, with integration along loop $C$ with $S$ being the area encircled by $C$.
Analogously, we take the right-hand-side of Eq.~(\ref{cbfb}) is something corresponds to Eq.~(\ref{eq21r}), and 
the left-hand-side to Eq.~(\ref{eq21l}).
Then, the relation
\begin{eqnarray}
 \partial_t \int_S {\bf B}\cdot d{\bf S}={h \over {2e}} {d \over {dt}} w_C[\chi]
 \label{eq22}
\end{eqnarray}
is obtained, where
${d \over {dt}} w_C[\chi]$ is given by
\begin{eqnarray}
 {d \over {dt}} w_C[\chi]={ 1 \over {2\pi}}\oint_C \partial_t (\nabla \chi )\cdot d{\bf r}-{ 1 \over {2\pi}}\bar{\bf v} 
 \times \int_S \nabla \times (\nabla \chi )\cdot d{\bf S}
 \label{eq22b}
\end{eqnarray}
Here, we choose $C$ to be so small that the above relation can be considered as a local one; and at the same time, it must be large enough so that the current and magnetic field can be treated as smooth ones. It is also assumed that an average value of ${\bf v}$ within $C$, $\bar{\bf v}$, can be taken as the flow velocity.

Recently, a revision of the standard superconductivity theory is presented  \cite{koizumi2022,koizumi2022b,Koizumi2023}. In this theory, the following is the condition for the superconducting state
\begin{eqnarray}
{d \over {dt}} w_C[\chi]=0, \quad w_C[\chi] \neq 0
\end{eqnarray}
This means that the supercurrent is a collection of loop currents that are protected by the topological winding numbers.
 Then, the condition in Eq.~(\ref{eq22}) becomes
\begin{eqnarray}
 \partial_t \int_S {\bf B}\cdot d{\bf S}=0
 \label{eq25}
\end{eqnarray}
This simply implies the local magnetic flux is stationary in the superconducting state.

\begin{figure}[H]
\begin{center}
\includegraphics[width=3.0cm]{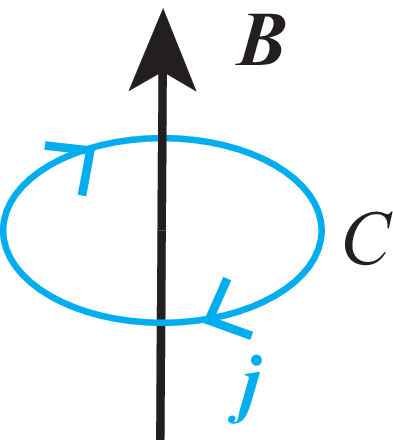}
\end{center}
\caption{A diamagnetic loop current ${\bf j}$ for the magnetic field ${\bf B}$. When ${\bf j}$ is a supercurrent, it is generated by $\nabla \chi$ with non-zero winding number $ w_C[\chi]$. In this case $\nabla \times (\nabla \chi) \neq 0$ at some points within the loop $C$.}
\label{DiaC}
\end{figure}

At a superconducting-normal state transition point in the phase diagram, however, 
\begin{eqnarray}
{ d \over {dt}} w_C[\chi] \neq 0
\end{eqnarray}
occurs. It should occur with keeping the condition in Eq.~(\ref{eq22}).
In order to interpret Eq.~(\ref{eq22}) with fluctuating $w_C[\chi]$, we consider the situation depicted in Fig.~\ref{DiaC}. 
For the $ w_C[\chi] < 0 $ case, the loop current is a diamagnetic one according to Eqs.~(\ref{eq12}) and (\ref{supercurrentd}); 
thus, the relation Eq.~(\ref{eq22}) means the increase of the magnetic flux is accompanied by decrease of the quantized diamagnetic loop current, and vice versa. In other words, the loop current considered here is an eddy current that is protected by the topological winding number.
The appearance and disappearance of this current is an abrupt one requiring the change of the integral number $w_C[\chi]$. Therefore, we may treat it as a discrete quantum transition. This quantum transition enables the reversible
superconducting-normal phase transition in a magnetic field, as will be discussed, below.

\section{The energy balance for superconducting-normal phase transition in a magnetic field: Moving phase boundary case.}
\label{Sec4}

Now we consider the situation where the interface at $x=0$ moves,
and the superconducting to normal phase change occurs in the interface region $0 <x <\lambda_L$.
Let us introduce the effective vector potential given by
\begin{eqnarray}
{\bf A}^{\rm eff}={\bf A}+{\bf A}^{\rm fic}
\end{eqnarray}
where 
\begin{eqnarray}
{\bf A}^{\rm fic}=-{\hbar \over {2e}}\nabla \chi
\end{eqnarray}
We treat the vector potential from the electromagnetic field and that from the Berry connection on equal footing,
and consider the sum of the two as the effective vector potential in materials.

Now we replace ${\bf E}$ in Eq.~(\ref{Faraday}) by
\begin{eqnarray}
{\bf E}^{\rm fic}=-\partial_t {\bf A}^{\rm fic}
\label{Eeff}
\end{eqnarray}
Then, we can adopt a particular solution to Eq.~(\ref{Faraday}) that avoids electric field generation, 
\begin{eqnarray}
\int_S \nabla \times {\bf E}^{\rm fic}\cdot d{\bf S}=-\partial_t \int_S {\bf B}\cdot d{\bf S}
 \label{eq11b}
\end{eqnarray}
This is rewritten as 
\begin{eqnarray}
\partial_t \int_S {\bf B}\cdot d{\bf S}={ h \over {2e}} \partial_t w_C[\chi]
\label{changeB-E}
\end{eqnarray}
where $\partial_t w_C[\chi]$ is defined as
\begin{eqnarray}
\partial_t w_C[\chi]={ 1 \over {2\pi}}\oint_C \partial_t (\nabla \chi )\cdot d{\bf r}
\end{eqnarray}
The relation in Eq.~(\ref{changeB-E}) is the one in Eq.~(\ref{eq22}) with neglecting the contribution from the flow term. 

The phase boundary shift occurs due to the offset of
the force balance in Eq.~(\ref{eqfb}); for example, if the Lorentz force exceeds the gradient of the kinetic energy force, the local velocity field stationarity in Eq.~(\ref{cbfb}) is violated; then, the quantized loop currents in the surface  region are replaced by the magnetic flux with keeping the condition in Eq.~(\ref{changeB-E}).
 This replacement is a direct one without generating the Joule heat. thus, enables the reversible
superconducting-normal phase transition.

A condition for the energy conserving conversion between the kinetic and magnetic field energies are
considered in Ref.~\cite{koizumi2020b}. The condition given there is Eq.~(10) of Ref.~\cite{koizumi2020b}, which is equivalent to Eq.~(\ref{Eeff}) (note that ${\bf E}$ and ${\bf B}$ in Ref.~\cite{koizumi2020b} respectively mean ${\bf E}^{\rm eff}$ and ${\bf B}^{\rm eff}$ in the present work). The relation in Eq.~(\ref{eq11b}) is the solution
for the condition in Ref.~\cite{koizumi2020b}.

Let us consider the energy balance problem from the total energy balance point of view.
Since the system is at a superconducting-normal state transition point in the phase diagram, the free energy density of the bulk of the superconducting region and that of the normal region (with including the magnetic field energy) is equal. 
Thus, the free energy change is equal to that in the interface region.
The magnetic field energy is given by
\begin{eqnarray}
F_m={ 1 \over {2 \mu_0}} \int d^3 r \ {\bf B}^2
\end{eqnarray}
and the kinetic energy is given by
\begin{eqnarray}
F_k={  m_e \over {2}} \int d^3 r \ {\bf v}^2 n_s 
\end{eqnarray}
and the sum of the two is calculated as 
\begin{eqnarray}
F_m+F_k=\left(\int dy dz \right) \lambda_L { 1 \over {2 \mu_0}} B_0^2
\end{eqnarray}
using ${\bf B}$ in Eq.~(\ref{lambda}) and  ${\bf v}$ in Eq.~(\ref{v1}),
where $(\int dy dz ) \lambda_L$ is the volume of the surface region.
It is equal to the energy of the normal phase magnetic field energy of the
volume $(\int dy dz ) \lambda_L$.
Thus, no loss of the energy occurs during the superconducting-normal phase transition.

At this point, we would like to respond to a claim presented in Ref.~\cite{Nikulov2021} since the statements there are
not correct. As explained above, what we need to explain the energy loss conversion of the superconductor-normal phase transition is the conservation of the $F_m+F_k$ in the surface region. This point is also explained in Ref.~\cite{Hirsch2017}, but missed in  Ref.~\cite{Nikulov2021}. The statement concerning Eq.~(4) in Ref.~\cite{Nikulov2021} is incorrect; as the non-zeroness of $w_C[\chi]$ in Eq.~(\ref{winding}) of the present work indicates,
\begin{eqnarray}
w_C[\chi]={1 \over {2\pi}}\oint_S \nabla \times (\nabla \chi) \cdot d{\bf S}\neq 0, 
\end{eqnarray}
which means that we cannot simply put $\nabla \times (\nabla \chi)=0$ in general (see Fig.~\ref{DiaC}). Actually, the same mistakes are seen in many textbooks, unfortunately.

\section{Concluding remarks}
\label{Sec5}

Using the velocity field given by Eq.~(\ref{eq12}), we have explained the reversibility of the
superconducting-normal phase transition in a magnetic field.
The key points are the appearance of the force term ${ m_e \over 2}\nabla v^2$ in Eq.~(\ref{eqfb}),
and the conversion of the quantized current to the magnetic flux given in Eq.~(\ref{changeB-E}).
For the appearance of ${ m_e \over 2}\nabla v^2$, treating the velocity of electrons as the velocity field is crucial.
The conversion given in Eq.~(\ref{changeB-E}) is beyond the Newtonian description of the dynamics.
It is expressed using the Faraday's electromotive formula type integrated interaction and the topological integer arising from the Berry connection.

Lastly, we would like to mention similarities between the present theory and the molecular vortex theory of Maxwell \cite{Maxwell1,Maxwell2,Maxwell3,Maxwell4}. Maxwell actually formulated electromagnetic equations using the vector potential, and considered the velocity field of the electric current. 
Later, his treatment was replaced by
the theory that removes the vector potential; and now four distilled equations (known as Maxwell's equations) and
the Lorentz force formula are considered to be the fundamental ones \cite{Lorentz} for electromagnetic phenomena.
After the advent of quantum mechanics and experimental verification of the Aharonov-Bohm effect, however, it is now established that the vector potential is a physical entity.  Further, the magnetic field arising from the singularities of the wave function considered by Dirac \cite{Monopole} seems to be rather general existence as the Berry connection \cite{Berry}; they give rise to the velocity field in the wave function formalism of quantum mechanics.
Therefore, 
the vector potential formalism employed by Maxwell seems to be more relevant than 
the standard one composed of four distilled Maxwell's equations and the Lorentz force formula. 
The elucidation of dynamical problems of electrons in materials in conjunction with the electromagnetic field will require the modernized version of the vector potential formalism put forward by Maxwell.

\section*{References}

\providecommand{\newblock}{}


\end{document}